\documentstyle[preprint,prl,aps]{revtex}
\tightenlines
\draft
\begin{document}
\title{Multiphoton Localization and Propagating Quantum Gap Solitons
in a Frequency Gap Medium}
\author{Sajeev John and Valery I. Rupasov\cite{pa}}
\address{Department of Physics, University of Toronto, Toronto,
Ontario, Canada M5S 1A7}
\date{June 26, 1997}
\maketitle
\begin{abstract}
The many-particle spectrum of an isotropic frequency gap
medium doped with impurity resonance atoms is studied using
the Bethe ansatz technique. The spectrum is shown to contain
pairs of quantum correlated ``gap excitations'' and their
heavy bound complexes (``gap solitons''), enabling the
propagation of quantum information within the classically
forbidden gap. In addition, multi-particle localization of
the radiation and the medium polarization occurs when such
a gap soliton is pinned to the impurity atom.
\end{abstract}
\vspace{7ex}


Light localization is a classical effect predicted \cite{J1}
to occur in strongly scattering dielectric microstructures.
In the context of photonic band gap (PBG) materials \cite{Y,J2},
nonclassical forms of localization such as photon-atom bound
states have been predicted \cite{JW} when the resonant transition
frequency of an impurity atom lies within a gap. This bound state
is an eigenstate of the quantum electrodynamic Hamiltonian for
a realistic PBG crystal exhibiting a general anisotropic photon
dispersion relation. In this state, a virtually emitted photon may
tunnel many wavelengths away from the atom before being reabsorbed,
leading to non-Markov memory effects \cite{JQ} in collective light
emission from many atoms. It was recently shown that an effective model
\cite{RS} describing both isotropic PBG systems and frequency
dispersive media (DM) \cite{LL} doped with resonance atoms exhibits
hidden integrability \cite{HI} and is diagonalized exactly \cite{RS,HI}
by means of the Bethe ansatz technique \cite{BA}.  This suggests
the possibility of a rich multi-particle spectrum in real physical
systems exhibiting a frequency gap, when such systems are doped
with impurity atoms.

In this paper, we demonstrate the existence of nonclassical states
of light which may be generated, for instance, through the interaction
of an external laser field with an impurity atom placed within
a polariton gap \cite{AM} of a DM. In addition to ordinary polaritons and
their bound complexes (``ordinary solitons'') occurring outside of the gap,
the sub-gap spectrum of the system is shown to contain propagating pairs
of correlated ``gap excitations'' and their heavy bound complexes
(``gap solitons''). The individual gap excitations comprising
the pair are correlated such that the probability amplitude of finding
them far apart decreases exponentially with the ratio of their separation
distance to the classical penetration length of the radiation into the
medium. In addition to heavy gap solitons propagating within the gap, the
spectrum contains multi-photon localized states pinned to the atom.
Under external perturbations a pinned gap soliton may dissociate into
propagating gap excitations. We evaluate the dispersion relations,
the effective masses, and the dissociation energies of quantum gap
solitons and we show that they are stable with respect to weak
perturbations. Our results demonstrate a clear distinction, at the
quantum level, between fermionic gap systems (such as electronic
semiconductors) and bosonic gap systems. In a semiconductor, propagation
within the energy gap is strictly forbidden. In the bosonic gap,
however, certain nonclassical many-body gap states are allowed to
propagate. This may have important consequences for the transmission
of quantum information within a bosonic gap medium. Although the Bethe
ansatz method made use of the {\em isotropic} one-particle dispersion
relation, qualitative similar results may hold in a realistic,
{\em anisotropic} PBG material which also exhibits a small
nonresonant Kerr nonlinearity \cite{JA}.

Consider the model Hamiltonian $\hat{H}=H_0+V$, where
\begin{mathletters}
\begin{eqnarray}
H_0 &=&\omega_{12}(\sigma^z+1/2) +\int_{C}\frac{d\omega}{2\pi}\,\omega
\,p^\dagger(\omega)p(\omega)\\
V&=&-\sqrt{\gamma}\int_{C}\frac{d\omega}{2\pi}z(\omega)[p(\omega)
\sigma^+ + p^\dagger(\omega)\sigma^-].
\end{eqnarray}
Here, $\omega_{12}$ is the transition frequency of the two-level
impurity atom placed within a bosonic frequency gap medium. The
spin operators $\vec\sigma=(\sigma^x,\sigma^y,\sigma^z)$,
$\sigma^\pm=\sigma ^x\pm i\sigma ^y$ satisfy the standard
commutation algebra $[\sigma^i,\sigma^j] =\epsilon_{ijk}\sigma^k$
and act on the atomic variables of the system.  The operators
$p^\dagger(\omega)$ ($p(\omega)$) create (annihilate) bosons of
frequency $\omega$ in a specific (electric dipole) spherical harmonic
state and satisfy the algebra
$[p(\omega),p^\dagger(\omega')]=2\pi\delta(\omega-\omega')$.
The integration contour consists of two parts, $C=C_-\oplus C_+$,
where $C_-=(0,\Omega_{\bot})$ and $C_+=(\Omega_{\|},\infty)$
correspond to the lower and upper branches of the medium excitations,
respectively.  The interaction term (1b) describes emission and
absorption of bosons by the atom. Here $\gamma=4\omega^3_{12}d^2/3$
is the inverse lifetime of the excited atom in free space with the dipole
transition moment $d$, while the atomic form factor $z(\omega)$ contains
the information about the polariton spectrum. Since the polariton gap
persists for small wave vectors, $k\rightarrow 0$, we can neglect
anisotropies of the underlying ionic crystal and take the dispersion
relation to be isotropic. Also, we choose units in which $\hbar=c=1$.

In the one-particle sector of the full Hilbert space, it is
straightforward to verify \cite{JW,RS,RS2} that when $\omega_{12}$ lies
inside the frequency gap, one of the eigenstates of the
Hamiltonian $\hat{H}$ describes a polariton-atom bound state.
The multi-particle sector of the Hilbert space can be studied by
the Bethe ansatz technique. The Schr\"odinger equation
$(\hat{H}-E)\vert\Psi_N\rangle=0$ is solved exactly \cite{RS,HI} due to
the two-polariton factorization of the multi-polariton scattering.
Imposing the periodic boundary conditions (PBC) on the $N$-polariton
wave function leads to the following set of Bethe ansatz equations (BAE):
\end{mathletters}
\begin{equation}
e^{ik_jL}\,\frac{h_j-i\beta/2}{h_j+i\beta/2}=
-\prod_{l=1}^{N}\frac{h_j-h_l-i\beta}{h_j-h_l+i\beta}\,,
\end{equation}
which completely determine the $N$-particle spectrum of the model (1).
Here, $E=\sum_{j=1}^{N}\omega_j$ is the eigenenergy and $\omega_j$ are
polariton frequencies which solve Eqs. (2).  Also $L$ is the radius of
a sphere centered at the atom on which we apply PBC and then take the
limit $L\rightarrow \infty$. In the case of a DM, the polariton momenta,
$k_j\equiv k(\omega_j)$, and ``rapidities'', $h_j\equiv h(\omega_j)$,
are expressed as,
\begin{equation}
k(\omega)=\omega n(\omega),\;\;\;
h(\omega)=\left(\frac{\omega_{12}}{\omega}\right)^2
\frac{\omega-\bar{\omega}_{12}}{\omega n^5(\omega)},
\end{equation}
with the refractive index $n(\omega)=\sqrt{\varepsilon(\omega)}$
and the dielectric permeability of the medium
$\varepsilon(\omega)=(\omega^2-\Omega^2_\|)/(\omega^2-\Omega^2_{\bot})$.
Here $\bar\omega_{12}\simeq\omega_{12}$ is the Lamb shifted atomic
transition frequency. The parameter $\beta=\gamma/\omega_{12}$
appears in both the polariton-atom scattering (l.h.s. of Eqs. (2))
and in the effective polariton-polariton coupling (r.h.s. of Eqs. (2))
caused by the polariton-atom scattering.

As $L\rightarrow\infty$, apart from real solutions, Eqs. (2) admit
complex ones, in which the rapidities $h_j$ are grouped into the
Bethe ``strings''. In this paper, we confine ourselves to the case
when all $N$ rapidities are grouped into a single string
$h_j=H+i\frac{\beta}{2}(N+1-2j),\;\;\;j=1,\ldots,N$
with a common real part (``carrying'' rapidity) $H$. This is a
solution of BAE if and only if the imaginary parts of rapidities
$h_j$ and corresponding momenta $k_j$ have the same sign:
\begin{equation}
{\rm sgn}({\rm Im}\,h_j)={\rm sgn}({\rm
Im}\,k_j),\;\;\;j=1,\ldots,N\ .
\end{equation}
This restricts possible magnitudes of polariton frequencies
$\omega_j$ corresponding to the string rapidities. It is easy to
understand that the necessary condition (NC) (4) determines
the frequency intervals, in which the effective polariton-polariton
coupling is attractive leading to bound many-particle complexes
(quantum solitons).

In empty space, an effective photon-photon coupling is attractive
for all frequencies of physical interest, and a Bethe string in the
space of rapidities is mapped to a quantum soliton in the space of
frequencies \cite{RY},
$
\omega_j=\Omega+i\frac{\gamma}{2}(N+1-2j)
$
with $E=\Omega N$. This also has a string structure. To avoid a possible
confusion in what follows we use the term ``string'' for solutions of
BAE in the $h$-space and the term ``soliton'' to refer to string's
images in the $\omega$- and $k$-spaces.

In the medium, analytical continuations of the functions $k(\omega)$
and $h(\omega)$ in the complex $\omega$-plane depend essentially on
the  position of the real part of the frequency with respect to the
medium gap. We start with the case when the real part of $\omega$
lies outside the gap. Let $\omega=\lambda+i\eta$ and the real part
$\lambda\in C$. For $\eta \ll \lambda$, the functions $k(\omega)$
and $h(\omega)$ are then represented as
$k(\omega)= k(\lambda)+i\eta k'(\lambda)$ and $h(\omega)=
h(\lambda)+i\eta h'(\lambda)$.
Since $k'(\lambda)\equiv dk(\lambda)/d\lambda$ is positive for all
$\lambda$, NC leads now to the condition $h'(\lambda)\equiv
dh(\lambda)/d\lambda>0$. This is met only if $\lambda\in C_-$.
Therefore, the effective coupling is attractive only between polaritons
of the lower branch, and the soliton image of the Bethe string
is given by $\omega_j=\Omega+\frac{i}{2}\Gamma(\Omega)(N+1-2j)$ and
$k_j=K(\Omega)+\frac{i}{2}Q(\Omega)(N+1-2j)$. Here
$K(\Omega)=\Omega n(\Omega)$, $\Gamma(\Omega)=\beta/h'(\Omega)$,
$Q(\Omega)=\beta k'(\Omega)/h'(\Omega)$, and $E=\Omega N$ is the
soliton eigenenergy. The common real part of the polariton frequencies
is found from the equation $h(\Omega)=H$, which has a root lying
in $C_-$ only if $H<0$. The soliton obtained is quite similar
to a vacuum soliton, and we will use the phrase ``ordinary soliton''
to refer to this solution, despite its inordinate behavior on different
polariton branches. Polaritons of the upper branch are described
by one-particle Bethe strings with real positive rapidities and
do not form any bound complexes. The results obtained are clearly
valid if $\omega_{12}\in C_-\oplus G$, where $G=(\Omega_\bot,\Omega_\|)$.
If $\omega_{12}$ lies above the gap, $\omega_{12}\in C_+$,
the effective coupling, including interbranch one, becomes attractive
and admits both ordinary solitons in each branch and unusual
``composite solitons'' containing polaritons of different branches.

Now let us look for an image of a Bethe string, provided the real
parts of all the frequencies $\omega_j$ lie inside the gap.
Let $\omega=\xi+i\eta$ and $\xi\in G$. To find the analytical
continuations of the functions $k(\omega)$ and $h(\omega)$ to the complex
$\omega$ space, we need first to fix an appropriate branch of the function
$n(\omega)$. Let $n(\xi\pm i0)=\pm i\nu(\xi)$, where
$\nu(\xi)=\sqrt{|\varepsilon(\xi)|}$. In this case
$k(\omega)={\rm sgn}(\eta)[-\eta\kappa^\prime(\xi)+i\kappa(\xi)]$,
where $\kappa(\xi)=\xi\nu(\xi)$. Also
$h(\omega)={\rm sgn}(\eta)[\eta\phi^\prime(\xi)-i\phi(\xi)]$
where $\phi(\xi)=(\xi-\omega_{12})f(\xi)$, and
$f(\xi)=\omega^2_{12}[\xi^3\nu^5(\xi)]^{-1}$.
Since the function $\kappa(\xi)$ is positive, NC leads to the
condition $\phi(\xi)<0$. It means that allowed gap excitations
exist only for $\xi\in(\Omega_\bot,\omega_{12})$. Because of a
strong nonradiative relaxation in the medium in the vicinity of
the frequency $\Omega_\bot$, we focus our studies on gap states
of physical interest lying in the vicinity of the atomic frequency
$\omega_{12}$. The remaining analysis is simplified by linearizing
the function $\phi(\xi)$ at the point $\xi=\omega_{12}$,
$\phi(\xi)\simeq a(\xi-\omega_{12})$, where $a=f(\omega_{12})$.

Now we are able to map a Bethe string to corresponding gap
excitations. We start with the simplest case of a two-particle string,
$N=2$. Its complex conjugated rapidities, $h=H+i\beta/2$ and
$h^*=H-i\beta/2$, are mapped to the corresponding pairs of the
complex conjugated frequencies, $\omega=\xi+i\eta$ and
$\omega^*=\xi-i\eta$, where the imaginary part is assumed to be positive,
$\eta>0$, and momenta, $k=q+i\kappa(\xi)$ and $k^*=q-i\kappa(\xi)$,
where $q=-\eta\kappa'(\xi)$. The real and imaginary parts of frequencies
are expressed in terms of the string parameters,
$\xi=\omega_{12}-\beta/2a$ and $\eta=H/a$. In the spherical
harmonic formalism introduced previously \cite{RS}, the real part of the
particle momenta $q$ must be positive. Since $\kappa'(\xi)$ is negative,
it follows that $q=\eta|\kappa'|$. Consequently only a string with
a positive carrying rapidity, $H>0$, is mapped to gap states of physical
interest.

The expressions obtained describe a novel, quantum correlated state
of two gap excitations. Two gap particles comprising the pair
are ``confined'' to travel together. They do not exist separately
from each other, unlike polaritons of ordinary solitons. Moreover,
the confined state cannot be treated as a bound state of two polaritons
from different branches (like a Wannier-Mott exciton in semiconductors),
because, under the condition $\omega_{12}\in G$, the interbranch
polariton-polariton coupling is repulsive. The spatial size of a pair,
$\delta\sim\kappa^{-1}(\xi)$, is nothing but the penetration length
of the classical radiation field with the frequency $\omega=\xi\in G$
into the medium \cite{LL}. Since the wave function of a single gap
particle is unnormalizable, free one-particle gap states in the bosonic
gap are forbidden in exactly the same way that electronic propagation
is forbidden in a conventional semiconductor gap.  However, in the
case of bosons, the effective particle-particle coupling allows
one to construct the normalizable wave function of a pair from
unnormalizable wave functions of each particle. At large interparticle
separations, the wave function of a pair has the form
\begin{equation}
\Psi(x_1,x_2)\sim\exp{\left\{iq(x_1+x_2)-\kappa(\xi)|x_1-x_2|
\right\}},
\end{equation}
where the real and imaginary parts of momenta describe, respectively,
the motion of the center of gravity and the spatial size of a pair.
The auxiliary coordinate variables $x_j$ are analogous to spatial
coordinates along an arbitrary axis passing through the atom.
The vicinities $x<0$ and $x>0$ correspond, respectively, to ingoing
and outgoing spherical harmonics of the polariton field. An angular
distribution of the field is determined by the specific spherical
harmonic polariton state.

Let us consider now the mapping of a string containing an even
number of particles (``even string''), $N=2l$, to gap excitations. A pair
of complex conjugated rapidities, $h_j$ and $h^*_j$, $j=1,\ldots,l$,
is mapped to a pair of frequencies $\omega_j=\xi^{(0)}_j+i\eta$ and
$\omega^*_j=\xi^{(0)}_j-i\eta$ and corresponding momenta
$k_j=q_j+i\kappa(\xi^{(0)}_j)$ and $k^*_j=q_j-i\kappa(\xi^{(0)}_j)$,
where the real parts of frequencies are given by
$\xi^{(0)}_j=\omega_{12}-(\beta/a)(l+1/2-j)$. The upper index, $(0)$,
indicates that these expressions are derived within the linear
approximation for $\phi(\xi)$ in the vicinity of $\xi =\omega_{12}$.
The expressions obtained describe a bound complex of $l$ pairs of
confined gap particles with the eigenenergy
\begin{equation}
E^{(0)}_l=2\sum_{j=1}^{l}\xi^{(0)}_j=2\omega_{12}l-(\beta/a)l^2\ .
\end{equation}
We will use the phrase ``gap soliton'' to refer to this state
of the system. Unlike an ordinary soliton, a gap soliton is stable
with respect to quite weak perturbations of the system, because the
dissociation energy, $U_d$, of a soliton with $l$ pairs into two
solitons with $l_1$ and $l_2=l-l_1$ pairs is positive:
$U_d\equiv
E^{(0)}_{l_1}+E^{(0)}_{l_2}-E^{(0)}_{l}=2(\beta/a)l_1l_2$.
The radial thickness of the spherical harmonic soliton pulse is
determined by the imaginary part of the momentum $k_1$ corresponding
to the rapidity $h_1$. Since $\kappa(\xi)$ is monotonically decreasing
function, the gap soliton size,
$\delta^{(0)}_l\simeq\kappa^{-1}(\xi^{(0)}_1)$, falls with the growth
of the number of pairs $l$.

Since the effective coupling constant is very small, $\beta\ll 1$,
the linear approximation works well even for large solitons containing
many pairs. But, in this approximation, the soliton energy is independent
of its momentum. For what follows it is convenient to introduce the soliton
energy per particle
$\epsilon^{(0)}_l\equiv E^{(0)}_l/2l=\omega_{12}-(\beta/2a)l$,
and the soliton momentum per particle,
$
q\equiv(2l)^{-1}\sum_{j}q_j\approx(\eta/l)\sum_{j}|\kappa'(\xi^{(0)}_j)|.
$
To estimate the first corrections to Eq. (6), we have to keep the next term,
$i(\eta^2/2)\phi''(\xi)$, of the Taylor series for the function
$h(\omega)$ and the next term in the expansion of the function
$\phi(\xi)$ at the point $\xi=\omega_{12}$:
$\phi(\xi)\simeq a(\xi-\omega_{12})+b(\xi-\omega_{12})^2$,
where $b=f'(\omega_{12})>0$. The frequencies $\xi_j$ are then given
by $\xi_j-\xi^{(0)}_j=-(b/a)(\xi^{(0)}_j-\omega_{12})^2+(b/a)\eta^2$,
while the soliton momentum is still given by $q$.
The term, $-(b/a)(\xi_j^{(0)}-\omega_{12})^2$, leads to the first
order correction to the energy of a motionless soliton and determines the
width of the soliton band, while $(b/a)\eta^2$ leads to the kinetic
energy contribution to the total soliton energy in the effective mass
approximation:
\begin{equation}
\epsilon_l=\epsilon_l^{(0)}-\Delta_l+q^2/2m_l.
\end{equation}
Here $\Delta_l=\frac{b\beta^2}{12a^3}(4l^2-1)$ and
$m_l=\frac{a}{2bl^2}\left(\sum_{j}|\kappa'(\xi^{(0)}_j)|\right)^2$ are
the band half-width and the effective mass, which increase with $l$.
At small $l$, the propagating gap soliton bands are very narrow and
solitons are very heavy and even motionless at $q=0$.
But the bandwidth increases as $l^2$, so that large $l$ solitons
are quite mobile when the momentum $q$ becomes larger than
the range of validity of the effective mass approximation. At arbitrary
$q$, the exact equations for the soliton parameters $\xi_j$ and
$\eta_j$ are given by ${\rm Re}\,h(\xi_j,\eta_j)=H$ and
${\rm Im}\,h(\xi_j,\eta_j)=\beta(l-j+1/2)$. The solution of these
equations requires simple numerical calculations.

Finally, we evaluate the pinning energy of a gap soliton to
the atom. In $h$-space, pinned solitons are described by
odd strings with $H\rightarrow 0^+$. The one-particle string,
$l=0$, with $H\rightarrow 0^+$ is clearly mapped to the gap state with
$\omega=\bar{\omega}_{12}\approx\omega_{12}$. This state is nothing
but the polariton-atom bound state \cite{JW,RS,RS2} in the one-particle
spectrum of the system. Therefore the extra real rapidity of an odd
string can be mapped to the gap state with $\xi^{(p)}_0=\omega_{12}$,
while  the remaining complex conjugated pairs of rapidities are mapped to
a deformed motionless gap soliton with
$\xi^{(p)}_j=\omega_{12}-(\beta/a)(l-j+1)$. We used here the term
``deformed'' to emphasize that the soliton frequencies now contain
the extra term $-\beta/2a$ due to the polariton-atom bound state,
which deforms the soliton, in contrast to the mobile, even gap
soliton whose frequencies are given by $\xi^{(0)}_j$.
Since one of the particles of an odd motionless gap soliton is
bound to the atom, a soliton as a whole is also ``pinned'' to the atom.
Moreover, as $H\rightarrow 0^+$, the imaginary parts of frequencies and
the real parts of particle momenta vanish. A pinned soliton describes
a many-particle state of the system, in which the radiation and medium
polarization are localized in the vicinity of the atom. To evaluate
the energy of pinning, we need only to compare the energy of a soliton
with $2l+1$ particles pinned to the atom,
$E^{(p)}_{l}=\omega_{12}(2l+1)-(\beta/a)l(l+1)$,
with the sum of the energies of the one-particle bound state,
$\omega_{12}$,  and a motionless gap soliton with $l$ pairs,
$E^{(0)}_l$. We find that the binding energy $U_l=-(\beta /a)l$  is
proportional to the number of pairs. Moreover, the energy required
to pull a single pair of particles out of a pinned soliton,
$U_1=(E^{(p)}_{l-1}+E^{(0)}_{1})-E^{(p)}_{l}=(\beta/a)(2l-1)$,
is even greater than the energy required to pull off all $l$
pairs. Therefore, the state of a pinned soliton is stable with
respect to quite weak perturbations of the system and its stability
increases with $l$. We mention finally that, for $H>0$, it is also
possible to construct composite solitons, consisting of an odd
number of bosons, which correspond, physically, to a bound state of
a gap soliton and a polariton in the upper branch.

In summary, we have shown that the isotropic dispersive medium
doped with an impurity atom exhibits a rich many-particle spectrum
containing heavy, mobile, gap solitons as well as pinned solitons.
Mobile gap solitons are highly nonclassical, quantum correlated
states consisting of an even number of gap particles.
This suggests the remarkable possibility that a bosonic frequency gap
medium, while impervious to classical linear wave propagation may
allow propagation of certain correlated quantum excitations.
Multi-particle gap solitons may be generated by both nonlinearly
exciting an impurity atom and Dicke superradiance from a collection
of these excited atoms. Unlike the single excitation which can tunnel
a distance given by the classical penetration length within the gap,
the paired excitations as well as the resulting heavy gap solitons can
propagate freely through the gap of this harmonic medium. The Bethe
ansatz solution which we have presented relied on the existence of
an isotropic polariton dispersion relation. In a real PBG material,
the photon dispersion relation is highly anisotropic. The propagation
of quantum information within a PBG material, in this manner, would be of
considerable importance in such applications as quantum computing \cite{BL}.

V. R. is grateful to the Department of Physics at the University
of Toronto for kind hospitality and support. This work was supported
in part by NSERC of Canada and the Ontario Laser and Lightwave Research
Centre.

\end{document}